\documentclass[12pt]{article}

\usepackage{amsmath,amssymb,amsfonts}
\usepackage[paper=letterpaper,margin=1.0in]{geometry}
\usepackage{graphicx}

\begin{document} 

\newcommand{\be}{\begin{equation}}
\newcommand{\ee}{\end{equation}}
\newcommand{\bea}{\begin{eqnarray}}
\newcommand{\eea}{\end{eqnarray}}
\newcommand{\ba}{\begin{array}}
\newcommand{\ea}{\end{array}}
\newcommand{\ben}{\begin{enumerate}}
\newcommand{\een}{\end{enumerate}}
\newcommand{\bi}{\begin{itemize}}
\newcommand{\ei}{\end{itemize}}
\newcommand{\bc}{\begin{center}}
\newcommand{\ec}{\end{center}}
\newcommand{\bfig}{\begin{figure}}
\newcommand{\efig}{\end{figure}}
\newcommand{\bq}{\begin{quotation}}
\newcommand{\eq}{\end{quotation}}
\newcommand{\bt}{\begin{table}}
\newcommand{\et}{\end{table}}
\newcommand{\btab}{\begin{tabular}}
\newcommand{\etab}{\end{tabular}}
\newcommand{\bs}{\begin{slide}}
\newcommand{\es}{\end{slide}}
\newcommand{\nn}{\nonumber}
\newcommand{\eref}[1]{(\ref{#1})}

\newcommand{\AdS}[1]{{\rm AdS}_{#1}}
\newcommand{\pa}{\partial}



\begin{center}
{\Large \bf Holographic Foam Cosmology: From the Late to the Early Universe}
\end{center}

\vskip 1.0cm

\centerline{{\bf
Y.\ Jack Ng
\footnote{\tt yjng@physics.unc.edu}
}}

\vskip 0.5cm

\begin{center}
{\it
Institute of Field Physics, Department of Physics and
Astronomy,\\
University of North Carolina, Chapel Hill, NC 27599, U.S.A.
}
\end{center}

\vspace{1cm}

\begin{abstract}

Quantum fluctuations endow spacetime with a foamy texture.  The degree of
foaminess is dictated by blackhole physics to
be of the holographic type.  Applied to cosmology, the holographic foam model
predicts the existence of dark energy with
critical energy density in the current (late) universe, the quanta 
of which obey 
infinite statistics. Furthermore we use the deep similarities between 
turbulence and the spacetime
foam phase of strong quantum gravity to argue that the early universe
was in a turbulent regime when it underwent a brief cosmic inflation with a 
``graceful" transition to a laminar regime.  In this scenario, both the late and
the early cosmic accelerations have their origins in spacetime foam.

 
\end{abstract}

\vspace{1cm}



\newpage

\section{Introduction}

There are two cosmic accelerations that we are aware of: a brief inflationary
acceleration \cite{inflat} in the early universe and the present 
(“late” universe”) acceleration \cite{SNa}  that is attributed to dark energy. 
Normally they are treated independently
and separately; but there are also some works \cite{BG,others} that 
consider both
regimes of accelerated expansions. 
We think it is conceptually necessary and aesthetically
pleasing to trace both cosmic
accelerations to a common cause in fundamental theory.  Following Wheeler \cite{Wheeler}, we
believe that space is composed of
an ever-changing geometry and topology called spacetime foam and that the
foaminess is due to quantum fluctuations
of spacetime. We argue for a scenario in which spacetime foam is the origin of 
both cosmic accelerations.\\

The outline of this Letter is as follows.  We begin with a brief review of the
holographic spacetime foam model.
In section 2 we use the quantum uncertainty principle coupled with black-hole physics to
show that spacetime is indeed foamy
and the degree of foaminess is consistent with the holographic principle; we
further argue that there necessarily exists
a dark sector in the universe. In section 3 we apply the holographic spacetime
foam to cosmology (with the
corresponding cosmology called holographic foam cosmology (HFC) \cite{Arzano,holocos}) 
and argue for the existence of dark energy with
critical energy density in the present universe and its quanta obey infinite
statistics.  In section 4 we use the deep similarities between 
the physics of turbulence and the universal geometric properties 
of the holographic spacetime foam to heuristically argue that the early universe 
was in a turbulent phase during which the universe
underwent a brief cosmic inflationary
acceleration.  Section 5 contains our concluding remarks.\\

We will use the subscript “P” to denote Planck units (with, e.g.,  
$l_P \equiv (\hbar G /c^3)^{1/2} \sim 10^{-33}$cm being the Planck length.) And for 
simplicity, $\hbar$, $c$, and the Boltzmann
constant $k_B$ are often put equal to unity.\\

\section{Holographic Spacetime Foam}

One manifestation of spacetime fluctuations is in the induced uncertainties in
any distance measurement. 
Consider the following gedanken experiment \cite{SW}
to measure the distance $l$ between 
a clock at one point and a mirror at another.
By sending a
light signal from the clock to the mirror in a timing experiment, we can
determine the distance.
The quantum uncertainty in the positions of
the clock and the mirror introduces an inaccuracy $\delta l$.  
Let us concentrate on the clock (of mass $m$).  If it has a linear
spread $\delta l$ when the light signal leaves the clock, then its position
spread grows to $\delta l + \hbar l (mc \delta l)^{-1}$
when the light signal returns to the clock, with the minimum uncertainty at
$\delta l = (\hbar l/mc)^{1/2}$.  Hence one concludes that
$ \delta l^2 \gtrsim \frac{\hbar l}{mc}. $
One can supplement this requirement with a limit from
general relativity\cite{wigner},  viz.,  $\delta l$ 
must be larger than the Schwarzschild radius $Gm/c^2$ of
the clock, yielding
$ \delta l \gtrsim \frac{Gm}{c^2}$, \footnote{Henceforth we will neglect multiplicative 
constants of order unity.}
the product
of which with the bound from quantum fluctuations finally 
gives \cite{wigner,Karol}
\begin{equation}
\delta l \gtrsim (l l_P^2)^{1/3} = l_P \left(\frac{l}{l_P}\right)^{1/3}.
\label{nvd1}
\end{equation}

This bound on $\delta l$ can also be derived by the following method which 
provides additional valuable insights.  Consider a spherical volume of
radius $l$ over the amount of time $T = 2l/c$ it takes light to cross the
volume.  One way to map out
the geometry of this spacetime region \cite{llo04}  
is to fill the space with clocks,
exchanging signals with
other clocks and measuring the signals' times of arrival. This process of
mapping the geometry is a sort of computation; hence
the total number of operations
is bounded by the Margolus-Levitin theorem\cite{Lloyd},
which stipulates
that the rate of operations for any computer cannot exceed the amount of energy
$E$ that is available
for computation divided by $\pi \hbar/2$.  To avoid collapsing the region into
a black hole, the total mass $M$ of clocks
must be less than $l c^2 /2 G$, corresponding to the upper bound on energy density
\begin{equation}
\rho \sim \frac{l/G}{l^3} = (l l_P)^{-2}.
\label{eq:rho}
\end{equation}
Together, these two limits imply that the
total number of operations that can occur in a spatial
volume of radius $l$ for a time period $2 l/c$ is no greater than $\sim
(l/l_P)^2 $. 
(Here and henceforth we set $c=1=\hbar$.)
To maximize spatial resolution, each clock must tick only once during the entire
time period. 
And if we regard the operations partitioning the spacetime volume into ``cells",
then
on the average each cell occupies a spatial volume no less than $ \sim l^3 / (
l^2 / l_P^2) = l l_P^2 $, yielding an average
separation between neighboring cells no less than $l^{1/3} l_P^{2/3}$.  This
spatial separation is interpreted as the
average minimum uncertainty in the measurement of a distance $l$, that is,
$\delta l \gtrsim l^{1/3} l_P^{2/3}$,
in agreement with the result Eq.(\ref{nvd1}) obtained above. \\

We can now heuristically derive the holographic principle.
Since, on the average, each cell occupies
a spatial volume of $ (\delta l)^3 \lesssim l l_P^2$, a spatial region of size $l$ can 
contain no more
than $l^3/(l l_P^2) = (l/l_P)^2$ cells.  Thus this
spacetime foam model corresponds to the case of maximum number of bits of
information $l^2 /l_P^2$ in a spatial region
of size $l$, that is allowed by the holographic principle
\cite{wbhts,GPH}.  Accordingly, we will refer to this
spacetime foam model (corresponding to $\delta l \gtrsim l^{1/3} l_P^{2/3}$)
as the holographic spacetime foam model.\\

\section{From Spacetime Foam to Dark Energy}

As a corollary to the above discussion, we can now give a heuristic
argument \cite{llo04,Arzano,plb} 
on why the Universe cannot contain ordinary matter only.
Start by assuming the Universe (of size $l$) has only ordinary matter.
According to
the statistical mechanics for ordinary matter at temperature $T$, energy scales
as
$E \sim l^3 T^4$ and entropy goes as $S \sim l^3 T^3$.  Black hole physics can
be
invoked to require $E \lesssim \frac{l}{G} = \frac{l}{l_P^2}$.
Then it follows that the
entropy $S$ and hence also the number of bits $I$ (or the number of degrees of
freedom) on ordinary matter are bounded by
$\lesssim (l / l_P)^{3/2}$.  We can repeat verbatim
the argument given in section 2 to conclude that, if
only ordinary matter exists, $\delta l \gtrsim \left( \frac{l^3}{(l /
l_P)^{3/2}} \right) ^{1/3} =  l^{1/2} l_P^{1/2}$ which is much greater than
$l^{1/3} l_P^{2/3}$, the result found above from our analysis of the
the gedanken experiment and implied by the holographic principle. 
Thus, there must be other kinds of matter/energy with which the Universe can map
out its spacetime geometry to a finer spatial accuracy than is possible
with the use of only conventional ordinary matter.
We conclude that a dark sector necessarily exists in the Universe!\\

The above discussion leads to the prediction of dark energy.  To see that,
let us now generalize this discussion for a static spacetime region
with low spatial curvature to the case of the recent/present
universe by substituting $l$ by $1/H$, where $H$ is the Hubble parameter.
\cite{Arzano,plb}   Eq.(\ref{eq:rho})
yields the cosmic energy density $\rho \sim
  \left(\frac{H}{l_P}\right)^2 \sim (R_H l_P)^{-2} \sim 10^{-120} M_P^4$.
Next, recall that we have also shown that the Universe contains $I \sim (R_H/l_P)^2$ bits
of information ($\sim 10^{120}$ for the current epoch).\cite{Arzano}  Hence the
average energy carried by each of these bits or quanta is $\rho R_H^3/I \sim
R_H^{-1}$. These long-wavelength bits or ``particles'' (quanta of spacetime foam)
carry negligible kinetic energy. (Note:
Such long-wavelength quanta 
\footnote{Alternatively one can
interpret these quanta as constituents of dark
energy, contributing a more or less uniformly distributed cosmic energy density
and hence acting as a dynamical effective cosmological constant
$\Lambda \sim H^2.$}
can hardly be called particles. We will simply call
them ``particles" in quotation marks.)
Since pressure (energy density) is given by kinetic energy minus (plus)
potential energy, a negligible kinetic energy means that
the pressure of the unconventional energy is roughly equal to minus its
energy density, leading to accelerating cosmic expansion,
in agreement with observation \cite{SNa}.
This scenario is very similar to that of quintessence \cite{quint},
but it has its origin in the holographic spacetime foam. \cite{holocos,yjng05} \\

How do these long-wavelength quanta differ from ordinary particles?
Consider $N \sim (R_H/l_P)^2$ such ``particles'' 
in volume $V \sim R_H^3$ at $T \sim
R_H^{-1}$, the average energy carried by each ``particle".  If these ``particles" 
obey Boltzmann statistics,
the partition function $Z_N = (N!)^{-1} (V / \lambda^3)^N$ gives
the entropy of the system $S = N [ln (V / N \lambda^3) + 5/2]$,
with thermal wavelength $\lambda \sim T^{-1} \sim R_H$.
But then $V \sim \lambda^3$, so $S$ becomes negative unless $N \sim 1$
which is equally nonsensical.  A simple solution is to stipulate that
the $N$ inside the log in $S$, i.e, the Gibbs factor
$(N!)^{-1}$ in $Z_N$, must be absent.  (This means that
the N ``particles'' are distinguishable!)
Then the entropy is positive: $S = N[ln (V/ \lambda^3) + 3/2] \sim N$. Now,
the only known consistent statistics in greater than 2 space dimensions
without the Gibbs factor is the quantum Boltzmann statistics, also known as
infinite statistics. \cite{DHR,greenberg} 
Thus we conclude that the ``particles'' constituting dark energy
obey infinite statistics, rather than the familiar Fermi or Bose
statistics. \cite{plb,minic}
For completeness, let us list some of the properties of
infinite statistics \cite{DHR,greenberg}.
A Fock realization of infinite statistics
is given by $a_k a^{\dagger}_l = \delta_{k,l}$.
It is known that particles obeying infinite statistics are distinguishable, and 
importantly their theories are non-local. \cite{fredenhagen,greenberg}
(To be more precise, the fields
associated with infinite statistics are not local, neither in the sense
that their observables commute at spacelike separation nor in the sense
that their observables are pointlike functionals of the fields.)
Their quanta are extended (consistent with what we show above for dark energy).
The number operator and Hamiltonian, etc.,
are both nonlocal and nonpolynomial in the field operators.  
This property of non-locality will be useful later in the discussion of the early Universe. 
But we should note that
TCP theorem and cluster decomposition still hold; and quantum field
theories with infinite statistics remain unitary. \cite{greenberg}\\

\section{From Spacetime Foam and Turbulence to Cosmic Inflation}

So far we have applied HFC to the present and recent cosmic eras 
(with $\rho \sim 10^{-120} M_P^4$).  But what about the early
universe (with $\rho \sim 10^{-8} M_P^4$)?  
Actually the discussion in the preceding section has already given us
some helpful hints \cite{Arzano}
especially with respect to inflation \cite{inflat} in the early universe.  
For example: (1) The flatness problem is largely solved because,
according to HFC, the cosmic energy is of critical density.  (2) It is quite
possible that HFC provides sufficient density perturbation as the model contains
the essence of a k-essence model. \footnote{Furthermore, the horizon problem may also be 
solved
since spacetime foam physics is essentially quantum blackhole physics and thus
is closely related to wormhole physics which can be used \cite{HK}
to solve the horizon problem.}  Nevertheless one important aspect of the early Universe  
appears to be missing in HFC.  It is connected to the expectation (supported by
Wheeler's insight \cite{wh55}) that, due to quantum fluctuations, spacetime, when probed at
very small scales, as is the case for the early universe, will appear very
complicated —- something akin in complicity to a chaotic turbulent froth.  So,
was spacetime turbulent in the early universe?  To this question we now sketch a
positive response.\\

First let us show the
deep similarities between the problem of quantum gravity and
turbulence \cite{jej08}.
The connection between them can be traced to the role of diffeomorphism
symmetry in classical gravity and the volume preserving diffeomorphisms
of classical fluid dynamics.  In the case of irrotational fluids in
three spatial dimensions,
the equation for the fluctuations of the velocity potential can be written in a
geometric
form \cite{unr95} of a harmonic Laplace--Beltrami equation:
$\frac{1}{\sqrt{-g}} \partial_a( \sqrt{-g} g^{ab} \partial_b \varphi) = 0 ~$,
where the effective space time metric has the canonical ADM form
$ds^2 = \frac{\rho_0}{c} [ c^2 dt^2 - \delta_{ij}(dx^i - v^i dt)(dx^j - v^j
dt)]$, with $c$ being the sound velocity. 
In this expression for the metric, it is apparent that the velocity of the fluid
$v^i$ plays the role of the shift vector
$N^i$ in the canonical Dirac/ADM treatment of Einstein gravity:
$ds^2 = N^2 dt^2 - h_{ij} (dx^i + N^i dt) (dx^j + N^j dt)$. 
Hence in the fluid dynamics context, $N^i \rightarrow v^i$, and a fluctuation of
$v^i$ would imply a fluctuation of the shift vector (and hence 
a fluctuation of the spacetime metric) and vice versa.\\ 

Next, let us note that, in fully developed turbulence in three spatial dimensions, Kolmogorov 
scaling specifies the
behavior of $n$-point correlation functions of the fluid velocity.
The scaling \cite{k41} follows from the assumption of constant energy flux,
$\frac{v^2}{t} \sim \varepsilon$, where $v$ stands for the velocity field of the
flow, and the single length scale $\ell$ is given as
$\ell \sim v\cdot t$.  This implies that $ v \sim (\varepsilon\, \ell)^{1/3}~, $
consistent with the experimentally observed two-point function
$\langle v^i(\ell) v^j(0)\rangle \sim (\varepsilon\, \ell)^{2/3} \delta^{ij}$.
Now recall our discussion above on
distance fluctuations $\delta \ell \sim \ell^{1/3} \ell_P^{2/3}$,
and define the velocity as
$v \sim \frac{\delta \ell}{t_c} $,
since the natural characteristic time scale is $t_c \sim \frac{\ell_P}{c} $,
then it follows that
$v \sim c \big(\frac{\ell}{\ell_P}\big)^{1/3} $. It is now obvious that a
Kolmogorov-like scaling \cite{k41} in turbulence has been obtained.
This interpretation of the Kolmogorov scaling in the quantum
gravitational setting implies that the quantum
fluctuation phase of strong quantum gravity in the early Universe 
could be governed by turbulence. \\

 
The discussion above is for the case of very large Reynolds number 
$Re = Lv/\nu$, where $v$ is the velocity field, $L$ is a characteristic scale 
and $\nu$ is the kinematic viscosity which is given by the product of the mean free path 
$\tilde{l}$ and an effective velocity factor $\tilde{v}$.
But was $Re$ actually large enough in the early Universe to set off turbulence?  For the
purpose of comparison and illustration, let us recall that, in conventional
cosmology at time, say, $10^{-35}$ sec., 
$vL$ in the numerator of $Re$ is roughly given
by the product of $v \sim 10^{-2} c$ and  $L \sim 10^{8} l_P$. \cite{Gibson}
For HFC close to
Planckian time, the denominator of $Re$ is given by $\nu \sim c l_P$ since, in that
regime, momentum transport could only be due to Planckian dynamics.  For the
discussion to follow, let us note that, relatively speaking, the 
effective velocity factor $\tilde{v}$ in
$\nu$ is not that different from the $v$ factor in the numerator of $Re$. The onset
of turbulence was due to
the smallness of the length scale (which plays the role of an effective 
mean free path) in the denominator of $Re$, viz., 
$\tilde{l} = l_P << L \sim 10^{8} l_P$ that was mainly responsible for
yielding a large $Re$ in the (very) early Universe. \\

There remains one crucial hurdle to overcome. If, as we suggest, turbulence
in the early Universe was related to inflation, we have to confront
the “graceful” exit problem: How to get a small enough
$Re$ to transit to the laminar phase and to end inflation in 
the process?  It is here the nonlocality property enjoyed by 
the quanta of spacetime foam 
(due to the fact that they obey infinite statistics) came to the rescue since 
the length scale $\tilde{l}$ in $\nu$ would naturally and eventually extend to the order
of $L$ \footnote{Compare with the case of dark energy discussed in Section 3.}
yielding a small enough $Re$ to suppress turbulence and to 
naturally end inflation.\\

 
\section{Conclusion}

We have sketched a scenario in which both the late and the early cosmic
accelerations have a common origin and can be traced to spacetime foam. 
The case for dark energy in the current/recent (``late") universe was proposed before 
\cite{Arzano,plb},  while the case for cosmic inflation in the early Univerese is the 
main 
focus of this Letter.  One attractive feature of our present proposal is that the scheme is 
very economical, involving no arbitrary or fine-tuned parameters.  It is also natural in that 
inflation was inevitable as turbulence set off by the Planckian dynamics was inevitable in the 
early Universe.   The scheme also 
provides a rationale for why inflation lasted only briefly (say, $ \sim 10^{-33}$ sec.) as 
the turbulent phase was quickly terminated due to the nonlocal (extended) property of 
the quanta of spacetime foam. Of course it is important
to check if this scenario is supported by more quantitative arguments and 
calculations.  In passing we should also mention that
it will be of great interest to see if our scenario for inflation 
discussed above can mitigate or at least amelioriate some of the criticism 
\cite{PenStein} against the inflation paradigm.\\

We conclude with two observations the first of which involves the crucial question of whether
our approach yields enough e-folds of inflation to solve the myriad of cosmological problems.
The following heuristic argument would seem to say probably there were.  
Let us follow the folklore: at the end of 
inflation, the energy stored in the quanta of spacetime foam would be converted into 
hot ordinary particles (as well as dark matter).  Since 
the Grand Unification era is around$\sim 10^{-34}$ sec., which, as an order-of-magnitude 
estimate, should 
also mark the end of inflation, giving enough (say $\gtrsim 65$) e-folds of inflation. 
This argument may be strengthened, if, as has been proposed \cite{Ho},
quantum gravity can actually be the origin of (ordinary-) particle statistics, 
and that 
infinite statistics (the statistics obeyed by quanta of spacetime foam) is the underlying 
statistics.  In that case, ordinary particles that obey Bose or Fermi statistics are actually 
some sort of collective degrees of freedom of ``particles" of infinite statistics.  (See Ref. 
\cite{Del} for a discussion of such a construction.)  Thus, arguably, at the end of inflation, 
quanta 
of spacetime foam could be converted into ordinary particles as required. \\

We end this paper with our second observation.  Our  aesthetically pleasing scenario can be 
compared to a recent model \cite{BG}
of quintessential inflation based on the assumption that the slow roll
parameter has a Lorentzian form as a function of the number of e-folds. 
Its form corresponds to the vacuum energy 
both in the inflationary (with $\rho \sim 10^{-8} M_P^4$) and the dark
energy (with $\rho \sim 10^{-120} M_P^4$)
epochs which are treated symmetrically.  In this model the inflationary
scale is exponentially amplified while the dark energy scale is suppressed,
producing a curious cosmological see-saw mechanism.  In the present work, 
the two cosmic accelerations are also attributed to a single mechanism;
but they are related by some sort of turbulent-laminar duality (or 
chaotic-smooth complementarity).  
It would be interesting to see if our scheme can be approximated by an effective 
theory in which a similar ansatz for the slow roll parameter naturally arises as in the 
see-saw model \cite{BG}.
But a full investigation may require a non-perturbative treatment of
quantum gravity, involving a truly non-local field theory of ``particles" obeying infinite 
statistics. \footnote{We note that
when the see-saw model is realized in the context of a single scalar field, the
extracted potential of the scalar field is fairly complicated.  Can this complication be 
a reflection of non-perturbative quantum gravity involving non-local quanta of spacetime 
foam?}
Handling turbulence in such a context of 
quantum gravity may also prove to be challenging.\\


\vspace{1in}

\noindent
{\bf Acknowledgments}

I thank David Benisty and Eduardo Guendelman for a useful correspondence.
I am grateful to the Bahnson Fund  and the Kenan
Professors Research Fund of the
University of North Carolina at Chapel Hill for partial financial support.
\\

\end{document}